\documentclass[aps,pre,twocolumn,showpacs,preprintnumbers]{revtex4}

\usepackage{amsfonts}
\usepackage{amsmath}
\usepackage{fontenc}
\usepackage[latin1]{inputenc}
\usepackage{dcolumn}
\usepackage{bm}
\usepackage{latexsym}
\usepackage{graphicx}
\usepackage{bbding}

\usepackage{color}
\usepackage{ulem}

\newcommand{\eref}[1]{(\ref{#1})}

\begin{document}

\title{The diffusive transport of waves in a periodic waveguide}

\author{Felipe Barra}
\address{Departamento de F\'isica, Facultad de Ciencias F\'isicas y Matem\'aticas, Universidad de Chile, Casilla 487-3, Santiago Chile}
\author{Vincent Pagneux}
\address{Laboratoire d'Acoustique de l'Universit\'e de Maine, UMR CNRS 6613 \\Avenue Olivier Messiaen, 72085 Le Mans Cedex 9, France}
\author{Jaime Zu\~niga}
\address{Departamento de F\'isica, Facultad de Ciencias F\'isicas y Matem\'aticas, Universidad de Chile, Casilla 487-3, Santiago Chile}

\date{\today}

\begin{abstract} 
We study the propagation of waves in quasi-one-dimensional finite periodic systems whose classical (ray) dynamics is diffusive. By considering a random matrix model for a chain of $L$ identical chaotic cavities, we show that its average conductance as a function of $L$ displays an ohmic behavior even though the system has no disorder. This behavior, with an average conductance decay $N/L$, where $N$ is the number of propagating modes in the leads that connect the cavities, holds for $1\ll L \lesssim \sqrt{N}.$ After this regime, the average conductance saturates at a value of ${\mathcal O}(\sqrt{N})$ given by the average number of propagating Bloch modes $\langle N_B\rangle$ of the infinite chain. We also study the weak localization correction and conductance distribution, and characterize its behavior as the system undergoes the transition from diffusive to Bloch-ballistic.
These predictions are tested in a periodic cosine waveguide. 
\end{abstract}


\maketitle 

\section{Introduction}\label{intro}
Wave propagation in periodic media has been studied quantitatively at least from the advent of quantum mechanics and the quantum theory of solids. 
Bloch-Floquet theorem, the underlying theoretical tool, allows to identify the propagating and non-propagating 
states that form 
bands as functions of the quasi-momentum. The group velocity of the propagating waves is
given by the derivatives of the energy bands 
with respect to the quasi-momentum, thus explaining the {\it ballistic} character of the Bloch states~\cite{kittel}.
Another well studied subject is wave propagation in disordered systems~\cite{pingsheng}. 
Here, three regimes are usually recognized depending on the system size $L$. When $L$ is smaller than the mean free path $\hat{\ell}$ the propagation is in the ballistic regime, if $\hat{\ell}\ll L  \ll \xi$, with $\xi$ the localization length, the system is in the diffusive regime and if $\xi \ll L$ it is in the localized regime. In three dimensional systems $\xi$ could be either finite or infinite and the transition between these two regimes is called the Anderson transition. 
In waveguides (quasi-one-dimensional systems)~\cite{saenz} $\xi \sim N\hat{\ell}$ with $N$ the number of modes in the scattering leads, 
so the diffusive regime can be observed in the semiclassical limit. The conductance is a natural quantity~\cite{abrahams} to study these wave properties in electronic, optical and acoustical systems. In disordered systems, its scaling with $L$ is such that, in the ballistic regime it is independent of $L$, in the diffusive regime it scales as $1/L$ and in the localized regime it decreases exponentially with $L$. 
While the ballistic and diffusive regimes have been observed in electronic systems (for a review see~\cite{Beenakker2}), localization has been more elusive but recently~\cite{page} has been experimentally observed with acoustic waves.
 
Since the experimental realization of photonic and phononic crystals~\cite{review} many interesting properties of waves in periodic media have been found~\cite{subwavelength transmission}. In this work we will consider a new one, namely the existence of a diffusive regime for waves in periodic media, a property usually associated to disordered systems. In fact, it is not always appreciated that the diffusive regime is a semiclassical property of some chaotic systems and disorder is not essential for its appearance.

There exists a wealth of literature where the classical (ray) dynamics of particles in periodic billiards is studied in relation to transport processes. The Lorentz channel~\cite{Gasp} is probably the best known example because is one of the few cases where hyperbolicity, the mathematical expression of hard chaos, has been proved. The Lorentz channel consists in a quasi-one-dimensional region populated by hard wall disks placed regularly in a lattice. Particles travel freely except for the elastic collisions with the obstacles. If the geometry of the lattice is such that there are no trajectories allowed to travel infinitely without collisions, then an initial density of trajectories will spread such that its variance grows proportional to time $t$, i.e. it will exhibit normal diffusion. 
Otherwise, diffusion is anomalous~\cite{bleher} and the particle density variance grows as $t\log{t}$. In this work we always assume normal diffusive dynamics. The essential ingredient for this diffusive behavior is the chaotic dynamics of the particle in the billiard unit cell. Hence, a natural question to ask is to what extent this classical diffusive behavior appears in the wave transport properties of periodic systems, especially when we think in the contrast between the well known ballistic wave propagation of Bloch states and the diffusive character of the classical dynamics.  
One approach to this problem is the one considered in~\cite{Dittrich:1998p118} were they studied the
dependence of spectral statistics on the diffusion coefficient of a ring of $L$ identical chaotic cells. 
Another approach is to study the time evolution of a wave-packet~\cite{Dorfman}. Unfortunately, this is numerically difficult if we are interested in the semiclassical regime. One can simplify this task by considering model systems like the spatially extended multibacker map~\cite{Dorfman}, where the diffusive to ballistic transition was exhibited in the mean square displacement. 

In this work, we consider a time-independent approach based on the scattering matrix and focus on finite quasi-one-dimensional periodic systems. Besides solving the wave equation in a particular waveguide system, namely the cosine billiard, we use Random Matrix Theory (RMT), which has been successfully used to model the wave properties of chaotic cavities~\cite{blumel} and of disordered wires, and also to study spectral properties of extended systems~\cite{Dittrich:1998p118}. Here, we employ RMT to study scattering in chaotic periodic waveguides. Since the unit cell of the periodic billiard is an open chaotic cavity, its scattering matrix can be modeled by elements of the Dyson circular ensembles~\cite{blumel}. Using this matrix, the scattering matrix of the slab with $L$ identical cells is constructed, from where physical properties, like conductance, can be computed~\cite{Landauer}. Averaging this conductance over the appropriate Dyson ensemble of random matrices we can obtain a prediction for the average conductance of a periodic waveguide composed of generic chaotic cavities, which we verify in the cosine billiard.

The plan of the paper is the following.  In Sec.~\ref{waveguide.sec} we review the basic tools to analyze scattering in waveguides, introduce the so called cosine periodic billiard and a random matrix model for periodic waveguides. 
In Sec.~\ref{cond}, 
we show numerically in the random matrix model and in the cosine billiard that for system length $1\ll L \lesssim \sqrt{N}$ the average conductance behaves diffusively, i.e. as $N/(L+1)$ and at a length of the order $\sqrt{N}$, the so-called 
diffusive-Bloch ballistic transition occurs and the average conductance saturates to a constant value. In two different subsections, 
we analyze conductance fluctuations and weak localization correction as the systems undergoes the transition from diffusive to Bloch ballistic. 
Finally, in Sec.~\ref{conc} we offer some conclusions. 

\section{Waveguide system}\label{waveguide.sec}

\begin{figure}[t]
 \centering
 \includegraphics[width=1\columnwidth]{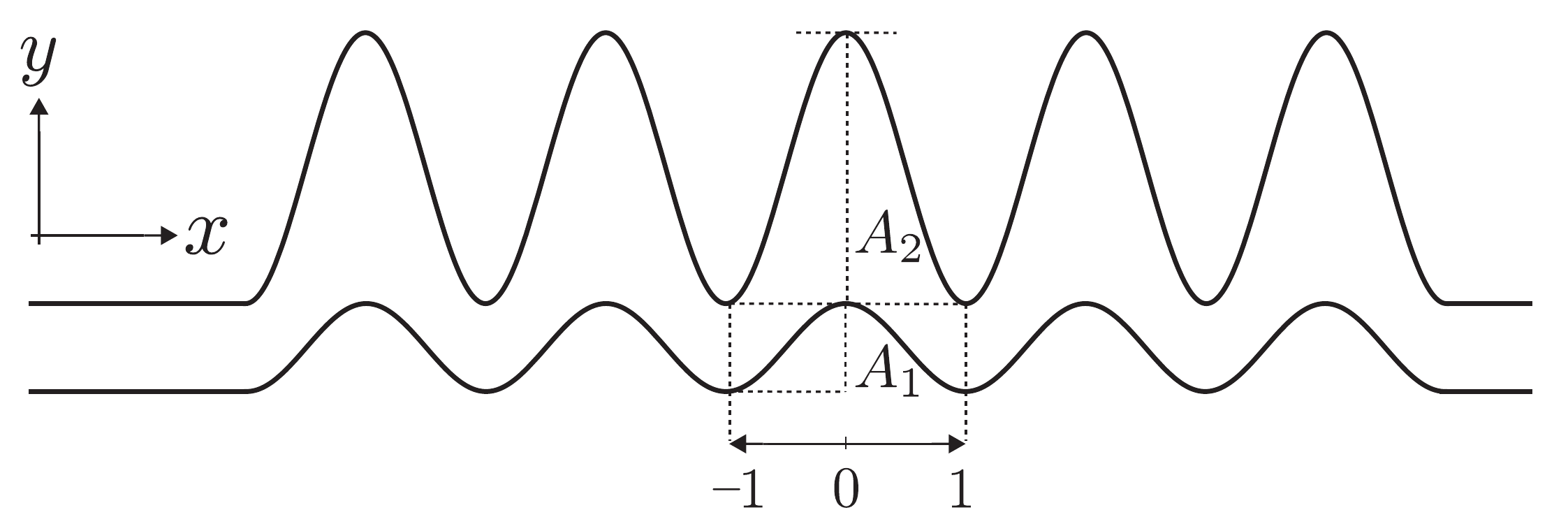}
 \caption
 {Cosine billiard chain with five unit cells connected to two plane leads. The unit cell boundaries are defined in \eref{cosine-billiard-def}--\eref{cosine-billiard-def-2} as a function of the amplitudes $A_1$ and $A_2$ shown in the figure. The width $W$ of the leads is $A_1$}
 \label{fig-cosine-billiard} 
\end{figure}

\begin{figure}
 \centering
\includegraphics[width=1\columnwidth]{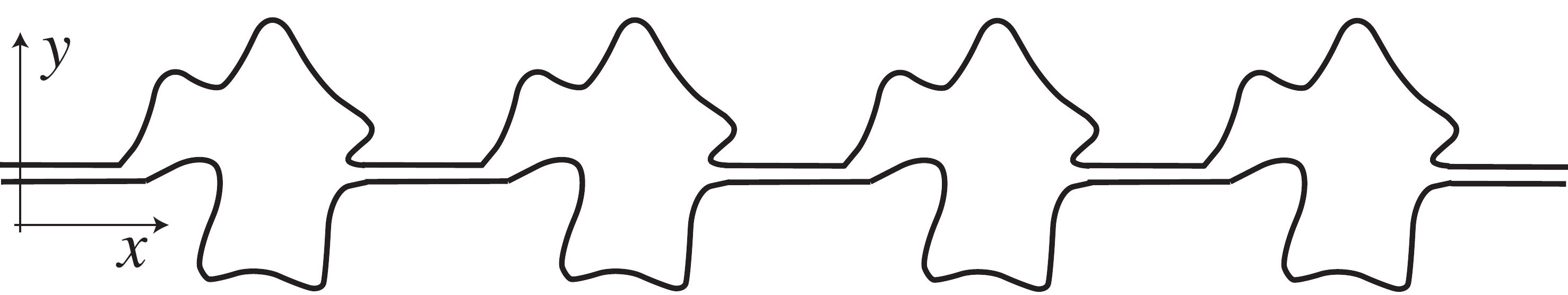}
 \caption{A schematic plot of a periodic chain of chaotic billiards. The leads that connect the chaotic cavities have a width $W$ such that
 if the wavenumber is $k$ there are $N=[W k/\pi]$ modes that propagates along them. The cavity has an area $A_c$ and the length of the unit cell is
 $a$ which we set to one unless we specify it otherwise. For the analysis of conductance the number of cells $L$ is finite (here $L=4$), the leftmost and rightmost leads extend to $x$ minus and plus infinity, and we deal with a scattering system. }
 \label{Fig1} 
\end{figure}

In this paper, the physical problem we address regards scattering in a waveguide 
composed of a slab made of a finite periodic chain of two dimensional chaotic cavities connected by leads [see Figs. \ref{fig-cosine-billiard} and \ref{Fig1}]. The leftmost and rightmost leads extend to $x$ going to minus and plus infinity.
The wave function $\phi$ is governed by the Helmholtz equation
\begin{equation}\label{helmholtz}
\nabla^2 \phi + k^2 \phi =0,
\end{equation}
with Dirichlet boundary conditions at the walls, and where $k$ is the wavenumber. Experimental realizations of this system can be built with microwave cavities~\cite{kuhl}. The number of cells in the slab is $L$ and we choose the $x$ direction as the waveguide axis.

A natural way to describe the wavefunction $\phi(x,y)$ in the waveguide is to project it on the local transverse basis, this is, writing the wavefunction as
\begin{equation}\label{LTM-general}
 \phi(x,y) = \sum_{n=1}^\infty{\left(c_n^{+}(x) + c_n^{-}(x)\right)\, \rho_n(x,y)},
\end{equation}
where $\rho_n(x,y)$ are the local transverse modes which satisfy the boundary conditions on each $x$, and $c^{+}_n(x)$ ($c^{-}_n(x)$) is the right-going (left-going) longitudinal mode. In the particular case of a hard-wall waveguide, 
\begin{equation}
 \rho_n(x,y)= \sqrt{\frac{2}{h(x)}}\,\sin{\left(n\pi \frac{y-h_1(x)}{h_2(x)-h_1(x)}\right)} ,
\end{equation}
$n=1,\ldots,\infty$, where $h_1(x) < h_2(x)$ are the walls height as a function of the longitudinal coordinate $x$. This set of functions satisfies the null boundary conditions $\rho_n(x,h_1(x)) = \rho_n(x,h_2(x))=0$ everywhere in the guide. The longitudinal modes are obtained by inserting (\ref{LTM-general}) in (\ref{helmholtz}), which transforms the original partial differential equation into a system of coupled ordinary differential equations which can be efficiently solved numerically~\cite{VincNum}. 

In a plane lead, $h_1(x)$ and $h_2(x)$ are constant, so (\ref{helmholtz}) is separable since $\rho_n(x,y)=\rho_n(y)$ is independent of $x$. In this region, the longitudinal modes $c^{\pm}(x)$ are given by 
\begin{equation}
e^{\pm}_n(x) = \frac{e^{\pm i k_n x}}{\sqrt{k_n}},
\end{equation}
where $k_n^2 = k^2 - \left(n\pi/W\right)^2$ is the longitudinal wavenumber, $W=h_2-h_1$ is the lead width and the normalization is to impose unit flux. In this region, there are 
\begin{equation}\label{N-open-modes}
N = \left\lfloor \frac{W k}{\pi} \right\rfloor
\end{equation}
propagating modes because for $n>N$ the longitudinal wavenumber $k_n$ is imaginary, implying null flux. These are called evanescent modes and decay exponentially with $x$. The far field wavefunction in the leads can be described with a $2 N$ dimensional complex vector composed of coefficients $A_n$ and $B_n$ for $n=1,\ldots,N$ and we can write the wave function as
\begin{equation}\label{LTM}
 \phi(x,y) = \sum_{n=1}^{N}{\left(A_n e^{+}_n(x) + B_n e^{-}_n(x)\right)\, \rho_n(x,y)} .
\end{equation}
Let $\bm A^{\rm r}$ ($\bm B^{\rm r}$) be the $N$ dimensional complex vector of right-going (left-going) amplitudes in the right lead and $\bm A^{\rm l}$ ($\bm B^{\rm l}$) the same on the left lead. We denote the incoming and outgoing (or incident and scattered) fields in vector notation as 
\begin{equation}
    \psi_{\rm in}=\left(\begin{array}{c}
   \bm A^{\rm l}\\
   \bm B^{\rm r}
 \end{array}\right) \quad \mbox{and}\quad
 \psi_{\rm out} =\left(\begin{array}{c}
   \bm B^{\rm l}\\
   \bm A^{\rm r}
 \end{array}\right).
\end{equation}
The scattering matrix $\bf S$ is defined as the linear transformation that maps incoming to outgoing fields,
\begin{equation}\label{s-matrix-def}
\psi_{\rm out} = \bm{S} \psi_{\rm in},
\end{equation}
and has a block structure,
\begin{equation}
\bm{S} =\left(
\begin{array}{cc}
\bm{r} & \bm{t'}\\
\bm{t} & \bm{r'}
\end{array}\right) 
\label{smatrix}
\end{equation}
with $\bm r$ the left ($\bm {r'}$ the right) reflection matrix and $\bm t$ the left to right ($\bm{t'}$ the right to left) transmission matrix, each of them of dimension $N\times N$. The scattering matrix of a chain of $L$ identical cavities (such as Fig.~\ref{Fig1}) can be obtained from a standard concatenation rule~\cite{prosen}, starting from the knowledge of the unit cell scattering matrix i.e. $\bm S_L = f_L(\bm S_{\rm uc})$ with $f_1(\bm S_{\rm uc}) = \bm S_{\rm uc}$, where $\bm S_{\rm uc}$ and $\bm S_L$ are the unit cell and $L$-cells scattering matrices, respectively. It is also useful to consider the transfer matrix $\bm M_L$, which maps the field on the left lead $\psi_{\rm l}$ to the field in the right lead $\psi_{\rm r}={\bm M_L} \psi_{\rm l}$, where
\begin{equation}
    \psi_{\rm r}=\left(\begin{array}{c}
   \bm A^{\rm r}\\
   \bm B^{\rm r}
 \end{array}\right) \quad \mbox{and}\quad
 \psi_{\rm l} =\left(\begin{array}{c}
   \bm B^{\rm l}\\
   \bm A^{\rm l}
 \end{array}\right).
\end{equation}
The concatenation rule for transfer matrices is a simple multiplication, so if $\bm M_{\rm uc}$ is the transfer matrix of the unit cell, then $\bm M_L=(\bm M_{\rm uc})^L$
and if $\lambda_i$ are the eigenvalues of $\bm M_{\rm uc}$  then $\lambda_i^L$ are the eigenvalues of $\bm M_L$. The consequence of this for the {\it infinite} 1D periodic systems are well known~\cite{cohen}. Since the wavefunction of the infinite periodic system must remain bounded along the chain, the only allowed states are those associated to the eigenvalues that satisfies $|\lambda_i|=1$. The number of these states is called the number of propagating Bloch states~\cite{Faure:2002p3} and will be denoted by $2N_B(k)$. Since in this case we can write $\lambda_i=e^{i\theta(k)}$, we can invert the relation and obtain the allowed \textit{energy} bands $k=k_n(\theta)$. 

As we have already mentioned, in this paper we are interested in the transport properties of {\it finite} periodic systems, in particular in the dimensionless conductance of a chain with $L$ cells, which can be obtained directly from the transmission part of the $\bm S_L$ matrix by the Landauer formula~\cite{Landauer},
\begin{equation}\label{landauer1}
g_k(L)={\rm tr}[\bm t_L {\bm t_L}^\dag].
\end{equation}

\subsection{Cosine waveguide}

As our particular model we will employ the periodic cosine billiard.  
We define the unit cell as the region enclosed by $h_1(x)<y<h_2(x)$ for each $x\in [-1,1]$, where 
\begin{eqnarray}
h_1(x) &=& \frac{A_1}{2}\left[1+\cos{(\pi x)}\right] \quad \mbox{and} \label{cosine-billiard-def}\\
h_2(x) &=& A_1+ \frac{A_2}{2}\left[1+\cos{(\pi x)}\right]. \label{cosine-billiard-def-2}
\end{eqnarray}
The classical limit of (\ref{helmholtz}) corresponds to noninteracting free particles within the system; 
collisions against the billiard boundaries $h_1$ and $h_2$ are elastic, thus the particle speed $v$ is constant. 
Note that our cosine billiard always has finite horizon, i.e. it does not allow unbounded collision-free trajectories for any values of $A_1>0$ and $A_2>0$, and is chaotic choosing these parameters appropriately~\cite{NbNous}. We always consider configurations displaying strongly chaotic dynamics such that classical particles in the cavity follow a normal diffusion process, {\it i.e.}
$\overline{x^2}\sim D t$, where the average $\overline{(\cdot)}$ is computed for each time $t$ over an initially spatially-bounded ensemble of initial conditions~\cite{NbNous} with random velocity.

We note that the unit cell mirror symmetry $x\to-x$ is not relevant for the classical transport properties of the billiard but makes the numerical solution of the quantum scattering problem faster (the transmission and reflexion matrices $\bm t$ and $\bm r$ are the same in both direction). However, this induces an anti-unitary symmetry in the quantum Hamiltonian which plays a role in the statistical and transport properties of the waveguide as  we discussed in~\cite{NbNous}.

We will solve numerically the scattering wave problem in this billiard as a function of $L$ and compute
the conductance \eref{landauer1}. In general, all quantities dependent on the $\bm M_{\rm uc}$ spectrum (in particular $g_k(L)$) are highly fluctuating as a function of $k$ over variations of the order of the (unit cell) mean level spacing~\cite{Dittrich:1998p118}. Averaging over a $k$ interval several mean level spacings wide gives a smooth function that changes on much larger $k$ scales. We call the ensemble of wavenumbers (or energies) realizations in this interval the \textit{semiclassical ensemble}. In the following sections, we will compute these averages for the cosine billiard and the results will be compared with the  predictions that follows from the RMT periodic waveguide model that we discuss below. If the typical size of the cavity is much larger than the leads width $W$, then it is possible to define the semiclassical ensemble over a wavenumber interval such that $N$ is constant. This assumption is important to make the connection to our RMT model.

\subsection{RMT periodic waveguide model}
The RMT model is constructed by taking a $2N\times 2N$ random matrix from the Dyson Circular Orthogonal (Unitary) Ensemble COE (CUE) and using it as the scattering matrix $S_{\rm uc}$ of a chaotic unit cell with (without) time reversal invariance. 
Modeling the scattering matrix of a chaotic cavity by elements of these ensembles is a common method~\cite{haake} known to be accurate in the prediction of statistical properties. 
The connection to the physical system is made by replacing $k$ dependent quantities by RMT realization dependent quantities. The averages over the RMT ensemble take the place of averages over the semiclassical ensemble. Using the $\bm S$ matrix composition rule $\bm S_L = f_L(\bm S_{\rm uc})$, the scattering matrix of the $L$ cells connected by leads with $N$ modes is obtained. Thus, we have a RMT ensemble for periodic chains of $L$ chaotic cavities.

If we denote by $\mu$ the Dyson measure of the RMT ensemble and note that the conductance \eref{landauer1} is a functional of $\bm S_L$, we can express the averaged conductance of the RMT model as
\begin{equation}
\langle g_N(L)\rangle=\int d\mu(\bm S_{\rm uc})g[\bm S_L(\bm S_{\rm uc})]
\end{equation}
We perform this computation numerically. 

To end this section we would like to remark that modeling a chaotic cavity by RMT
is justified if the particle stay trapped inside the cavity for a long time~\cite{haake}. To be more precise, the RMT model assumes that the particle escape time is much longer than its correlation decay characteristic time, so the particle effectively undergoes a random walk between unit cells. Therefore, the case of anomalous diffusion mentioned in the introduction is excluded from our analysis. In order to study this case, the Dyson ensembles should be properly modified.

\section{Conductance of a finite periodic chain of chaotic cavities}\label{cond}

Landauer's formula can be written more explicitly as
\begin{equation}
g_k(L)={\rm tr}[\bm t_{L} { \bm t_L}^\dag]=\sum_{i=1}^N T_i(L),
\label{landauer}
\end{equation}
where $\{ T_i(L) \}_{i=1}^N$ are the $N$ eigenvalues of the $N\times N$ matrix $\bm t_{L} { \bm t_L}^\dag$ which are bounded in the real interval [0,1]. They are related to the eigenvalues $\{\Lambda_i(L), \Lambda_i(L)^{-1}\}_{i=1}^{N}$
of the $2N\times 2N$ matrix $\bm M_L\bm M_L^\dag$ by 
\begin{equation}\label{polar-rel}
T_i(L) = \frac{4}{2 + \Lambda_i(L) + \Lambda_i^{-1}(L) } \,,\quad i=1,\ldots,N ,
\end{equation}
a relation that follows from the 
polar decomposition~\cite{BeenakkerRMP}. To simplify notation, we drop the explicitly $k$ dependence in all quantities below; we note that $N=[W k/\pi]$ is also $k$ dependent.

\begin{figure}[t]
 \centering
 \includegraphics[width=1\columnwidth]{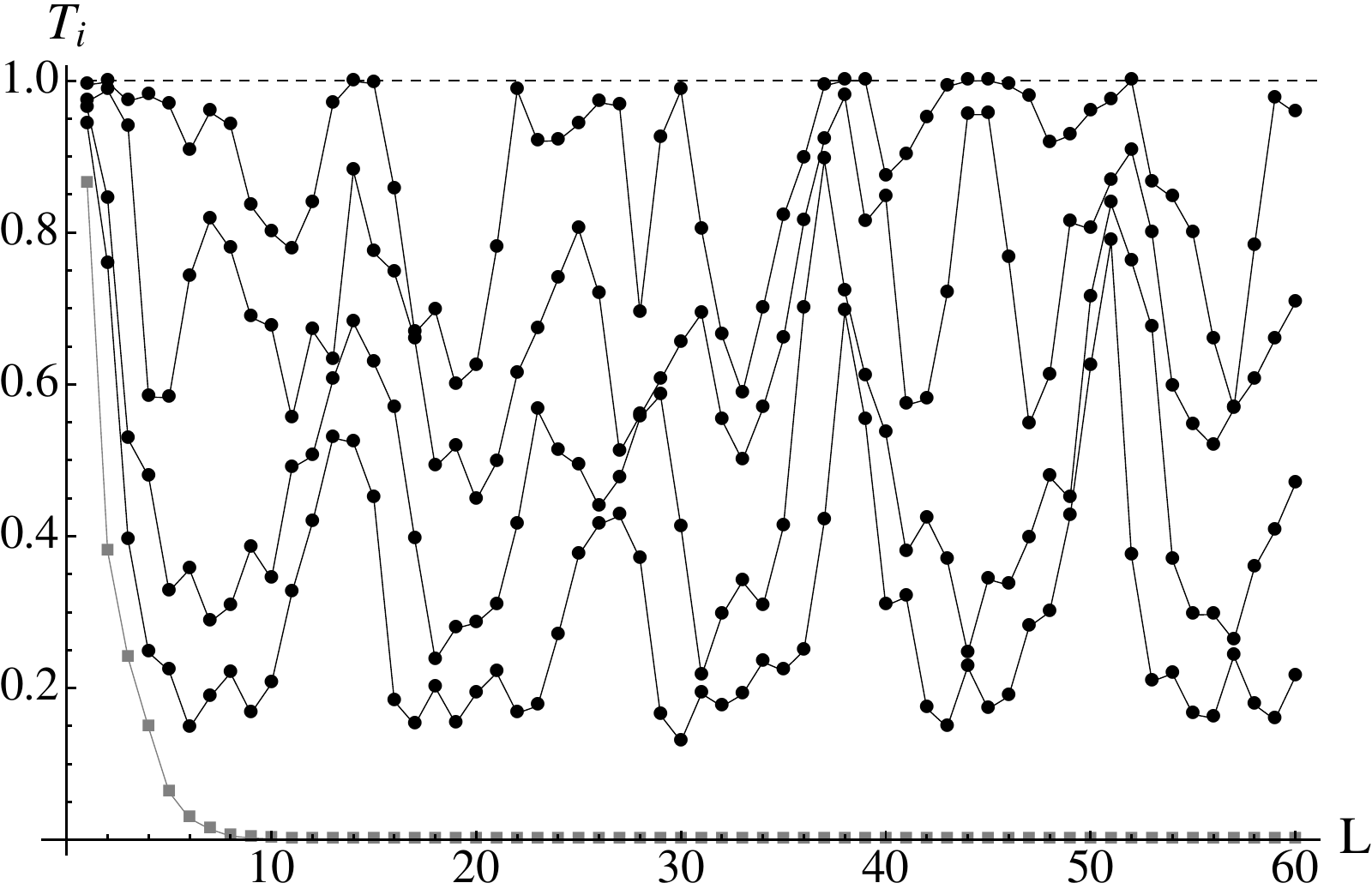}
 \caption {The first five transmission eigenvalues $T_i(L)$ in a cosine billiard with $A_1=0.5$ and $A_2=4.5$ for $k=30.2157\pi$. For this energy, $N_B$=4 and the associated $T_i$ are plotted with circles. In addition, the slowest to decay evanescent mode is also plotted with squares. It can be seen that the ballistic transmission modes tend to spread in the [0,1] interval and therefore $\langle g(L)\rangle < N_B$ in the generic case.  }
 \label{fig-T_i} 
\end{figure}

The conductance \eref{landauer} -- and in general any other transport property dependent of the transmission eigenvalues $T_i$ -- is a function of $\bm{M}_L\bm{M}_L^\dagger$ eigenvalues.
In view of the simple description of the infinite periodic system in terms of allowed and forbidden states it is interesting to link the eigenvalues $\{\Lambda_i(L), \Lambda_i(L)^{-1}\}_{i=1}^{N}$ to the eigenvalues $\lambda_i$ of $\bm M_{\rm uc}$. Oseledets theorem~\cite{oseledets1968} provides us such relation, given by~\cite{BeenakkerRMP}
\begin{equation}\label{oseledets} 
\Lambda_i(L) \underset{L\rightarrow\infty}{\longrightarrow} a_i(L) e^{-2L \log{|\lambda_i|}},
\end{equation}
where $a_i(L)$ is a positive and  (generically) bounded function of $L$. Then, using relation (\ref{polar-rel}), we can decompose $g(L)$ [Eq.(\ref{landauer})] in two terms, one with the sum of the $N_B$ non-decaying transmission modes $T_i$ related by Eqs.~(\ref{polar-rel}) and (\ref{oseledets}) to the $2 N_B$ propagating Bloch modes $|\lambda_i|=1$ (which we choose to have indices $i=1,\ldots,N_B$), and another with the sum of the transmission modes related to evanescent Bloch states $|\lambda_i| \neq 1$ (which we choose to have indices $i=N_B+1,\ldots,N$). The second term has a decay length 
\begin{equation}\label{T-decay-length}
\ell = \left(\min_{|\lambda_i|>1}\{\log|\lambda_i|\}\right)^{-1}  ,
\end{equation}
determined by the slowest to decay non-propagating state. From  (\ref{oseledets}) and  (\ref{polar-rel}) we deduce that the $N_B$ transmission eigenvalues $T_i$ associated to Bloch modes is of order one, so for chains of length $L\gtrsim \ell$,
\begin{equation}\label{g_long_L}
g(L) \lesssim N_B + 4 a_{m}(L)\, e^{-2L/\ell} ,
\end{equation}
where $a_m$ is the Oseledet function $a_i$ associated to the mode with decay length $\ell$. The equality in \eref{g_long_L} is non-generic and occurs if $\bm M_{\rm uc}$ is a \textit{normal} matrix (as we discuss below), in which case $T_i(L)=1$ for all Bloch modes. More generally, for chains of length $L\ll\ell$, all transmission modes contribute to the conductance, 
\begin{equation}
g(L)=N_B P(L)+\sum_{i=N_B+1}^N   \frac{4}{2 + \Lambda_i(L) + \Lambda_i^{-1}(L) } 
\label{g}
\end{equation}
with $0<P(L)<1$ an ${\mathcal O}(1)$ quasi-periodic function of $L$. The function $P(L)$ takes into account the fact that there exists a \textit{repulsion} between the eigenvalues $T_i(L)$ associated to propagative Bloch modes and that these quantities fluctuate quasi-periodically as a function of $L$ [see Fig.~\ref{fig-T_i}]. This effect makes the conductance strictly lower than $N_B$ in the generic case, and can be thought as a geometric consequence of $\bm M_{\rm uc}$ being not normal, i.e. not diagonalizable in a orthogonal base. The transmission eigenvalues repulsion is also observed in disordered chains~\cite{BeenakkerRMP}.  

We remark that $P(L)$, $N_B$ and the spectrum $\lambda_i$ are $k$ 
dependent. Intervals where $N_B(k)=0$ corresponds to forbidden bands and Eq.~(\ref{g}) shows, as expected, that for $k$ values in a forbidden band, $g(L)$ decrease exponentially with $L$, while in allowed bands, where $N_B(k)\geq 1$, the conductance remains finite as $L\to \infty$. As $k$ increases, the probability of finding an energy gap decreases, nevertheless gaps play an important role in some statistical measures [see Fig.~\ref{fig-1_g}].

\subsection{Digression: the Disordered chain}
\label{sec.3.a}

We will contrast our results for the periodic RMT chain with a disordered RMT chain, where the composition of the scattering matrix is performed each time with a different realization of the CUE or COE ensemble. The statistical properties of the conductance of this disordered chain are expected to be well described by the statistical properties of the disordered wire which allow an analytical study~\cite{BeenakkerRMP}. 
One of the theoretical frameworks to study this model is through a Fokker-Planck equation for the probability $P(T_1,\ldots,T_N;L)$. In this approach, the system size $L$ is a continuous 
variable and the transmission coefficients are random variables that evolves stochastically as $L$ increase. There are important hypothesis in the derivation of this Fokker-Plank equation (called DMPK equation in this context~\cite{dorokhov1982,dmpk}), for instance, the disorder is spatially homogeneous and the scattering produced by a short slab of the wire is isotropic and weak. From $P(T_1,\ldots,T_N;L)$ it is possible to obtain all the desired statistical information of the conductance, which can be written as $\hat{g}(L)=\int dT P_L(T)T$ with
$P_L(T)={\mathcal N}\int  dT_2\ldots dT_NP(T,T_2,\ldots,T_N;L)$ (with ${\mathcal N}$ a normalization constant ).
From the solution of the DMPK equation, it is possible to show that the conductance behaves diffusively i.e. 
as $\hat{g}(L)=N\hat{\ell}/(L+1)$ for $\hat{\ell}<L<\hat{\ell} N$, and then, for $L>\hat{\ell} N$, as $\hat{g}(L)\propto \exp(-L/2\beta N\hat{\ell})$ with $\beta=1$ ( $\beta=2$) for systems with (without) time reversal invariance and $\hat{\ell}$
the mean free path in the disordered medium up to a numerical constant~\cite{BeenakkerRMP}. From this analysis 
it also follows that the localization length for the disordered wire is $\hat{\ell} N$.
It is also possible to characterize the fluctuations for instance to give an explicit expression for $P_L(T)$. 
For instance $P_L(T)=\frac{N\hat{\ell}}{2L}\frac{1}{T\sqrt{1-T}}$, the so called Dorokhov distribution, in the diffusive regime $\hat{\ell}<L<\hat{\ell} N$. This result is linked to the distribution of the localization lengths spectrum $\hat{\ell}_n=\ln |\hat{\lambda}_n|$ defined from the eigenvalues of the disordered ${\bf M}$ matrix satisfying $|\hat{\lambda}_n|>1$. Acording to Dorokhov~\cite{dorokhov1982}  $\hat{\ell}_n(N)^{-1} \sim n/N\hat{\ell}$ for $n=1,\ldots,N$. 
It is interesting to note, that as a consequence of $\hat{\ell}_n^{-1}$ linear dependence on $n$, 
\eref{polar-rel} implies that $\hat{g}(L) \sim 1/L$ in the interval $\hat{\ell}\ll L\ll\hat{\ell}N$.
There are other two important results that follow from the DMPK equation. Firstly, there is a weak localization correction (WLC) that has to be added to $\hat{g}(L)$ in the metallic (ohmic) regime in time reversal invariant systems, which for large $N$ is $\delta \hat{g}(L) = -1/3$ independent of $L$. 
Secondly, the conductance distribution in the metallic regime turn out to be Gaussian  with ${\mathcal O}(1)$ variance, independent of the system length and of the disorder properties (universal conductance fluctuations), and then changes from this Gaussian form to a log-normal distribution as the size of the wire reaches the localization length.

Other approaches have been used to study the conductance of a disordered wire, for instance using a random matrix theory for the 
description of the hamiltonian in the scattering region~\cite{hans}, but in the appropriate limits of the weakly disordered wire these descriptions were shown to be equivalent. It is expected that the disordered RMT chain is well described by this model, except perhaps in the transition between the different regimes. Our numerical results [see Fig.\ref{Fig3} (lower panel)] confirm this expectation.   
For instance, a transition from the diffusive to the localized regime when the system reaches a length 
$L\sim{\mathcal O}(N)$ corresponding to the localization length.

\subsection{Average conductance of the periodic waveguide}

In this section we focus on the average conductance $\langle g(L) \rangle$ and resistance $\langle 1/g(L) \rangle$, that we compute in the semiclassical and RMT ensembles for a periodic waveguide. From the decomposition \eref{g} of $g(L)$ as a sum of propagating and decaying modes, it is clear that for small values of $L$ we have $\langle g(L)\rangle \sim {\mathcal O}(N)$, 
as in the disordered case~\cite{BeenakkerRMP} and, for $L$ large enough,
$\langle g(L)\rangle\lesssim \langle N_B\rangle$ is independent of $L$.
The average number of propagating Bloch modes
in the infinite periodic waveguide, $\langle N_B\rangle$, was studied in~\cite{Faure:2002p3} and~\cite{NbNous}, where it was shown that for the diffusive waveguides we consider it scales as $\langle N_B\rangle\sim \sqrt{N}$. Therefore, as $L$ increases, there is a transition from the ballistic propagation $\langle g(L)\rangle\sim {\mathcal O}(N)$ through short systems to the Bloch-ballistic propagation $\langle g(L)\rangle\lesssim \langle N_B\rangle\sim {\mathcal O}(\sqrt{N})$ through long finite periodic systems. In order to study this transition we compute $\langle g(L)\rangle$ as a function of $L$ for the cosine periodic waveguide and for the RMT model of a periodic waveguide. We focus in particular in the limit of large $N$, i.e. the semiclassical limit. 

\begin{figure}
\centering
\includegraphics[width=1\columnwidth]{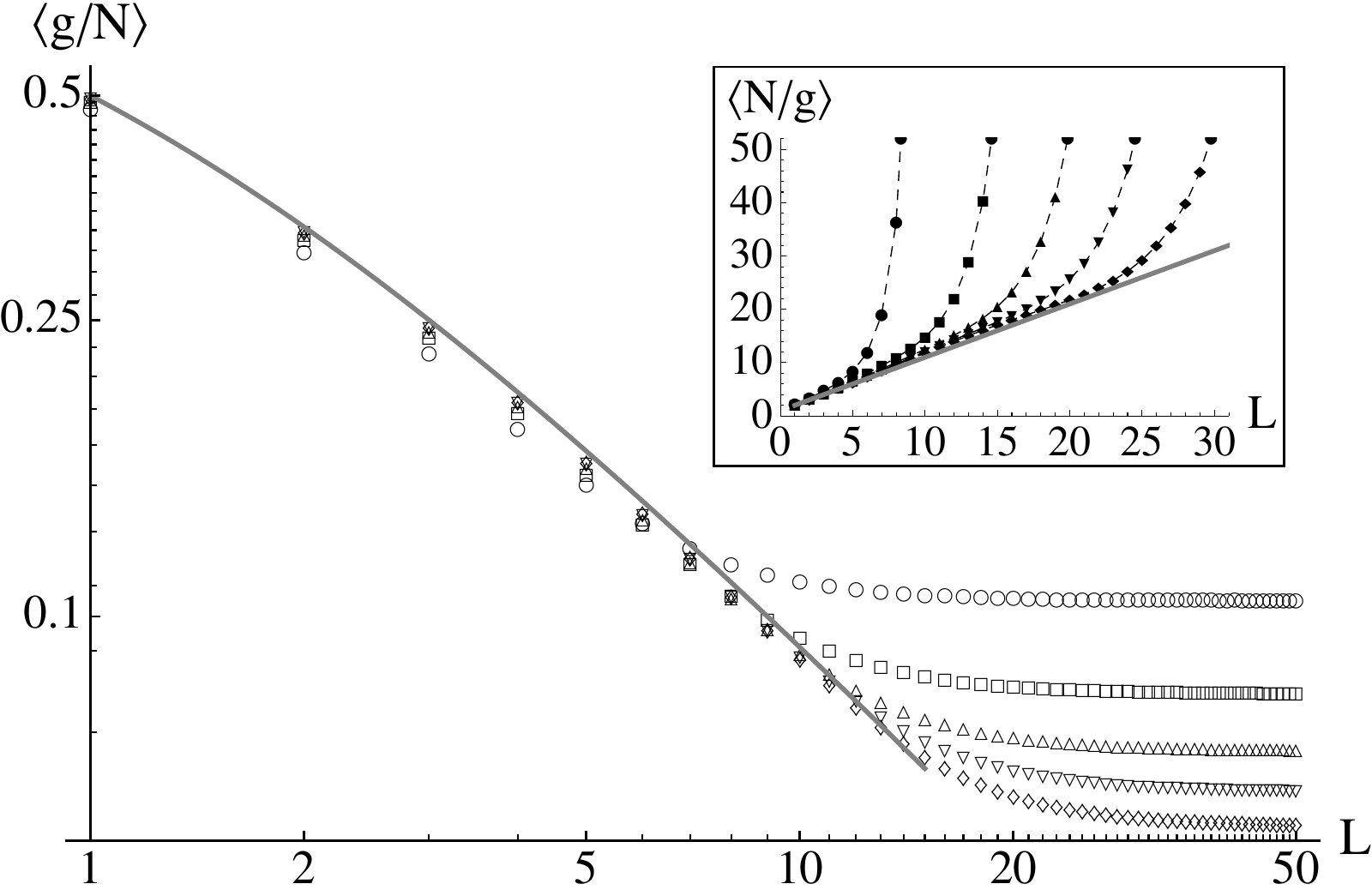}\\
\includegraphics[width=1\columnwidth]{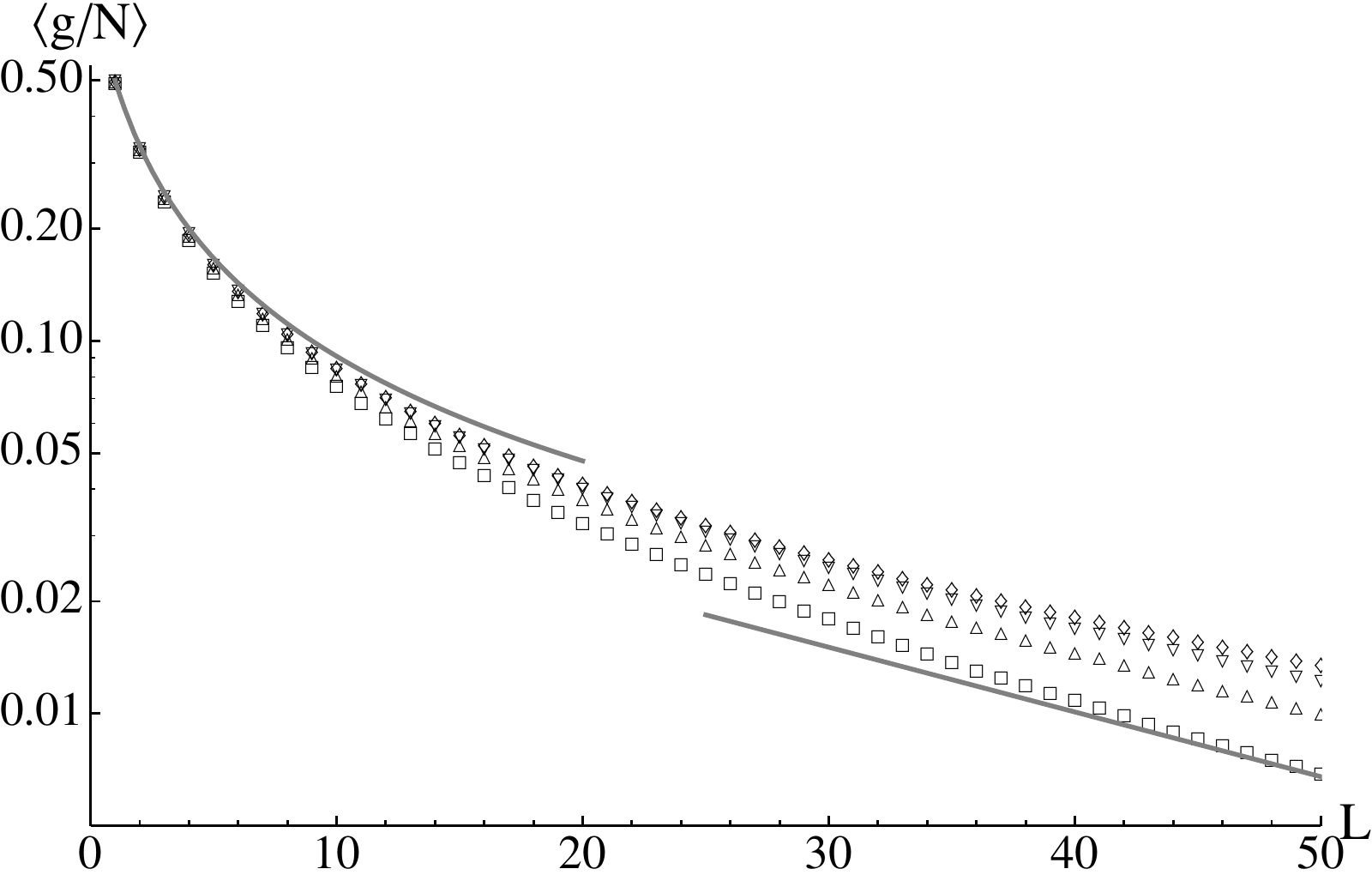}
\caption{Upper panel: For the periodic RMT chain, plots of $\langle g(L)/N \rangle$ as a function of $L$ for $N=10$ (dots), $20$ (squares), $30$ (triangles up), $40$ (triangles down) and $50$ (diamonds) for the COE case. As $N$ increase, convergence to the linear dependence on $1/(L+1)$ (line) and the scaling with $N$ of the diffusive regime are clearly observed. The departure from this law to an $L$ independent regime signals the diffusive to Bloch ballistic transition. 
In the inset we depict $\langle N/g(L)\rangle$ for the same values of $N$ and the line $L+1$ predicted by Ohm's law.
The divergence of this quantity that follows after the ohmic behavior is discussed in the text.
 Lower panel:
For the disordered RMT chain  (see sect. \ref{sec.3.a}), plots of  $\langle g(L)/N \rangle$. Symbols correspond to the same matrix dimension as in upper panel ($N=10$ not shown). The continuous line for $L<20$ represents Ohm's law and for $L>20$ an exponential decay as expected in the localized regime.}
\label{Fig3} 
\end{figure}

In Fig.~\ref{Fig3} (upper panel) we present the scaled averaged conductance $\langle g(L)/N \rangle$ as a function of $L$ for our RMT model with $\bm S_{\rm uc}$ taken from the COE ensemble and compare it 
with the characteristic $1/(L+1)$ decay of the ohmic behavior~\cite{saenz}. We observe that the numerical data follows 
\begin{equation}
\langle g(L)\rangle_{\rm COE}=\left\{
\begin{array}{cc}
\displaystyle\frac{N}{(L+1)} & 1 \ll L \lesssim \sqrt{N}\\
\displaystyle\langle N_BP(L) \rangle& \sqrt{N}  \ll L 
\end{array}\right.
\end{equation}
with the ohmic behavior,
$ \langle g(L) \rangle=N/(L+1)$,
up to a given length $\mathcal{O}(\sqrt{N})$ close to $\langle N_B\rangle$ where a transition to an $L$ independent regime is observed. 
This is the diffusive to Bloch-ballistic transition. 
\begin{figure}[h!]
 \centering
 \includegraphics[width=1\columnwidth]{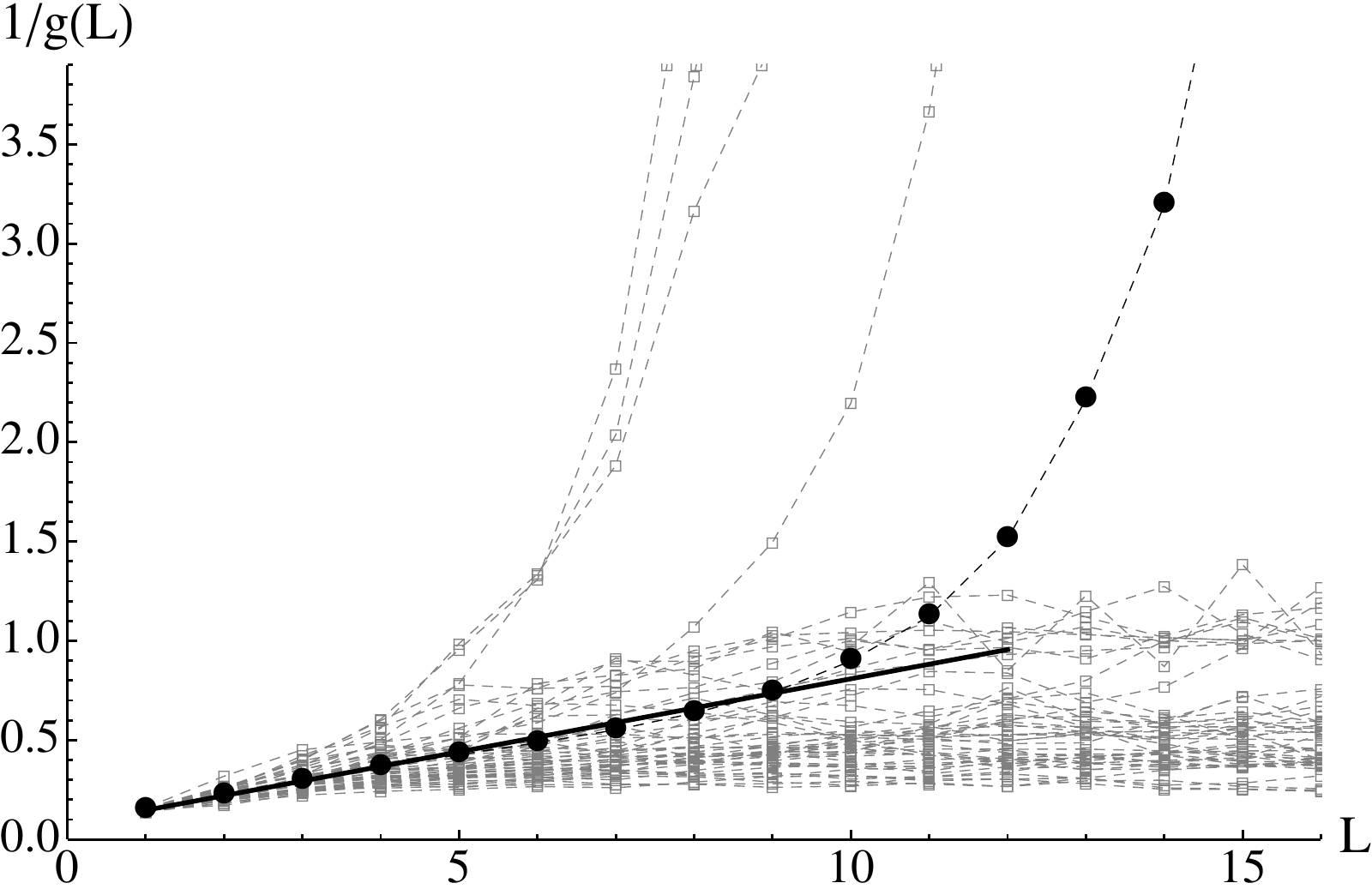}
 \caption {Average resistance $\langle 1/g(L)\rangle_k$ (black dots) for the cosine billiard with $A_1=0.5$, $A_2=4.5$ and $k=30.33\pi$. The full line shows the ohmic regime holding for $L<10$. Also, several realizations of $1/g(L)$ are displayed in gray: a few cases with $N_B=0$ grow exponentially and the rest reach an asymptotic oscillating value $1/g_\infty(L) \sim \mathcal{O}(1/N_B)$. The exponential grow of the average resistance $\langle R\rangle_k$ for $L>10$ is explained by the presence of the rare cases with $N_B=0$. }
 \label{fig-1_g} 
\end{figure}
The ohmic regime manifest itself also in the resistance,
\begin{equation}
\langle R \rangle = \left\langle \frac{1}{g(L)} \right\rangle,
\end{equation}
which is plotted in the inset of Fig.~\ref{Fig3} (upper panel) for the RMT model and in Fig.~\ref{fig-1_g}  for a cosine waveguide. Qualitatively, we found Ohm's law $\langle R\rangle = (L+1)/N$ holds for small $L$, before $\langle g(L)\rangle$ reaches its asymptotic value. This regime is followed by localization-like exponential grow of the resistance. The latter may seem surprising but is a consequence of the non-null probability of $N_B=0$. In fact, although this probability decays to zero as $N\to\infty$~\cite{tobe}, it dominates in the average $\langle 1/g(L) \rangle$ for long chains. However, in the limit of large $N$, for particular realizations of $1/g(L)$ the most probable is $1/g(L)=1/(N_B P(L))<\infty$ with $N_B\neq0$.

\begin{figure}[t]%
 \centering
 \includegraphics[width=1\columnwidth]{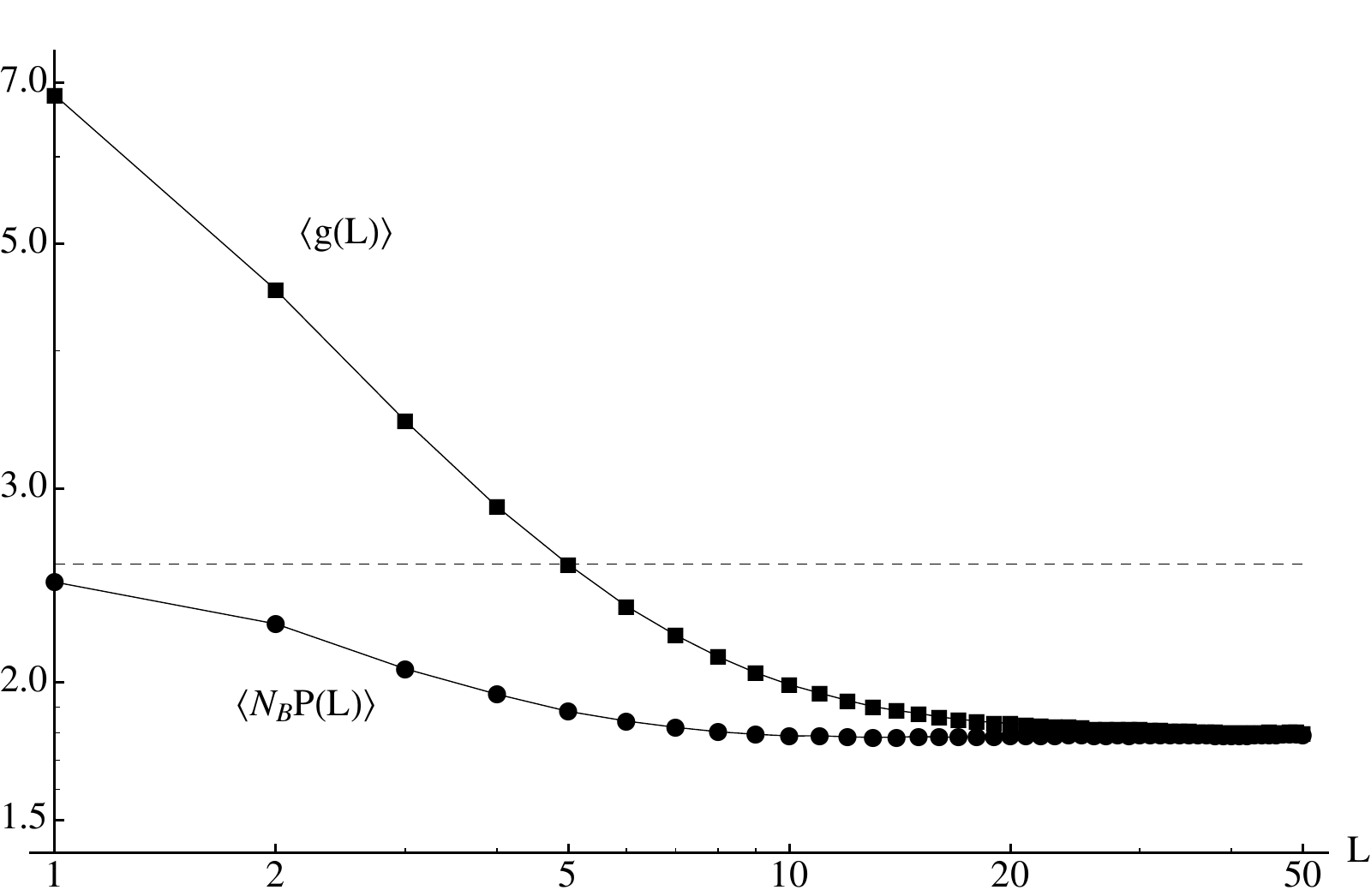}
 \caption
 {average conductance of the cosine billiard (filled circles) is compared to the average of (\ref{g}) eliminating the second term of the right hand side (filled squares).}
 \label{fignonprop} 
\end{figure}
\begin{figure}[t]
 \centering
 \includegraphics[width=1\columnwidth]{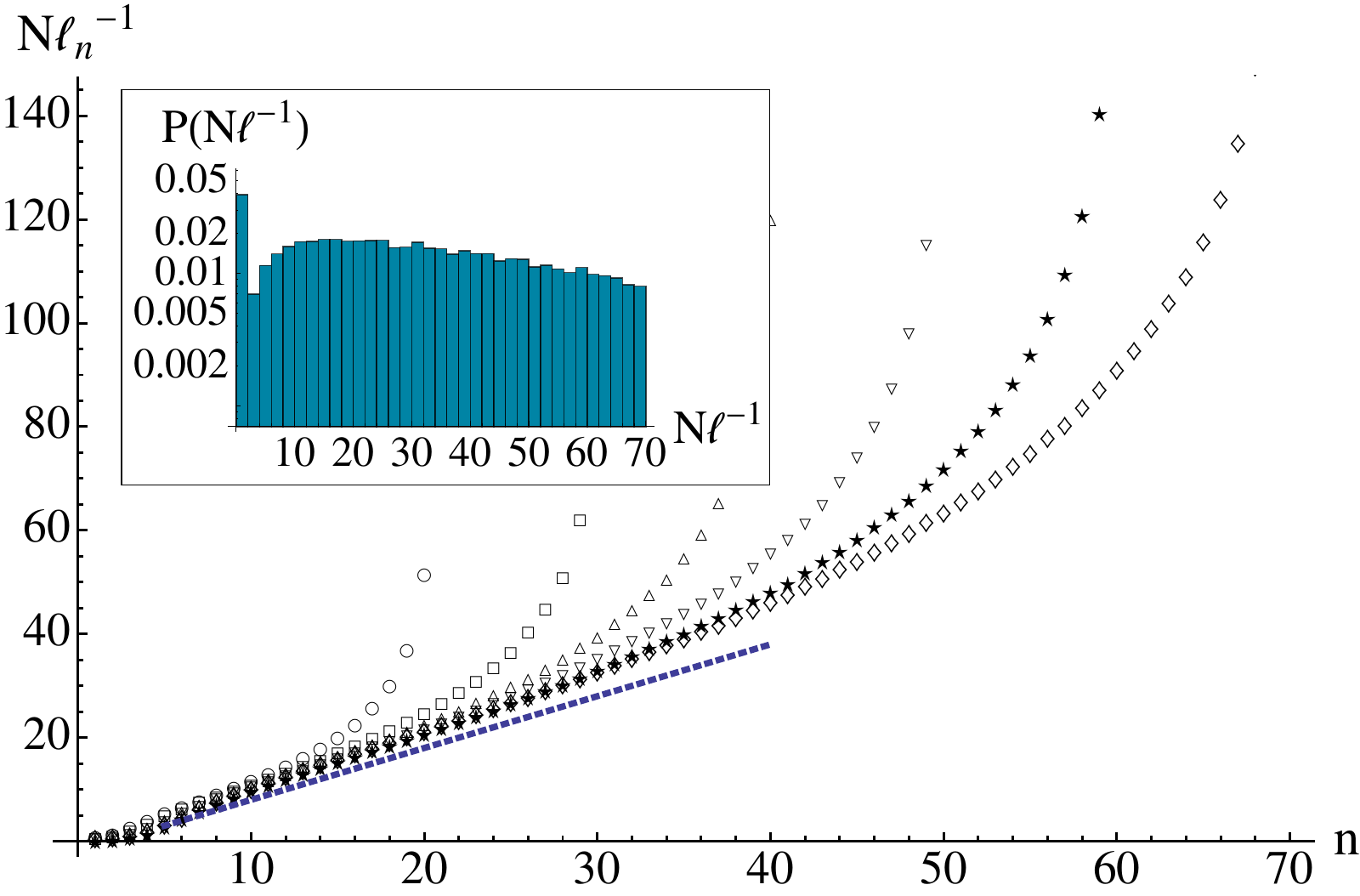}
 \caption
 {Average inverse decay lengths spectrum $\ell_n^{-1} = \log{|\lambda_n|}$ for the COE periodic chain model. The plot shows $N\ell_n^{-1}$ for $N=10,20,30,40,50,60,70$ (circles, squares, up triangles, down triangles, stars, diamonds). The integer $n$ is the absolute-value-sorted $\lambda_n$ eigenvalue index. For this system $\langle N_B\rangle = 4.7$. As can be seen in the plot $N\ell_n^{-1}$ is close to zero for $n\lesssim \langle N_B \rangle$ and is followed by a range with linear growth $N\ell_n^{-1} \sim n$ (dashed line) which explains the existence of the ohmic regime in the periodic chain for $L< \langle N_B \rangle/N$. In the inset, part of the pdf $P(N\ell^{-1})$ for $N=70$ is plotted, showing a constant range for $10\lesssim N\ell^{-1}\lesssim 30$ which is equivalent to $P(|\lambda|)\sim|\lambda|^{-1}$ [see \eref{lambda-conjecture}]. }
 \label{fig-coe-lambdas} 
\end{figure}

The existence of the ohmic regime is a consequence of the full $\bm{M_L M_L^\dagger}$ matrix spectrum and it is observed for for lengths $L\ll \ell$, where several non-propagating $T_i$ still contribute to the sum in the right hand side of (\ref{g}). 
This is clearly observed in Fig.~\ref{fignonprop} where the average conductance of the cosine billiard (filled squares) is compared to the average of (\ref{g}) eliminating the second term of the right hand side (filled circles).
In order to understand this ohmic regime we remember (see Sec.\ref{sec.3.a}) that, for the disordered wire,
it can be explained by the scaling of Dorokhov localization lengths spectrum~\cite{dorokhov1982}  
$\hat{\ell}_n(N)^{-1} \sim n/N\hat{\ell}$, $n=0,\ldots,N-1$. 
Now, it turns out that, in our model for periodic waveguides, the spectrum $|\lambda_n|$ of $\bm{M}_{\rm uc}=\bm{M}_{\rm uc}(\bm S_{\rm uc})$ with $\bm{S}_{\rm uc}$ taken from COE has a similar property, namely $\ell_n^{-1}=\log{|\lambda_n|} \approx n/N$ for $\langle N_B\rangle\leq n \ll N$, from which we conclude the existence of an ohmic regime, in periodic diffusive waveguides [see Fig.~\ref{fig-coe-lambdas}]. This is a statistical characteristic of the transfer matrix spectrum, which can be recasted stating that 
\begin{equation}\label{lambda-conjecture}
p(|\lambda|) \sim \frac{1}{|\lambda|} \;,\quad \mbox{for}\quad \frac{\langle N_B\rangle}{N} < \log{|\lambda|} \ll 1
\end{equation}
with $p(|\lambda|)$ the marginal pdf of the spectrum absolute values $|\lambda_n|$. From this follows that $1/g\sim L$ for $L \lesssim N/\langle N_B\rangle  \sim \sqrt{N}$, which can also be understood from the two following arguments. 

First, if we consider that the conductance in the diffusive regime $\langle g(L)\rangle \approx N/(1+L)$ must match its asymptotic average value $\langle g_\infty(L)\rangle \sim \langle N_B\rangle$, we obtain that this happens at $L\sim N/\langle N_B\rangle$. 

Second, if we consider a wave-packet representing an electron in a periodic system, we expect from semiclassical arguments to observe chaotic diffusion if many energy bands contribute to the superposition. This is the case if the diffusion time $t_D=L^2/D$ is smaller than the unit-cell
Heisenberg time $t_H = mA_c/\hbar$, where we recall that $D=D_1 v$, with $v$ the particle's speed and $m$ the electron mass. Then, we obtain that $t_D<t_H$ if and only if $L \lesssim \sqrt{A_c D_1 k} \sim \sqrt{N}$ which is the expected result. For longer times (correspondingly $L\gtrsim \sqrt{N}$) the energy bands are resolved and the wave-packet starts to propagate ballistically.  

\subsection{Weak localization}

Having established the existence of the diffusive regime we turn to the issue of weak localization. So far we have focused only on time-reversal symmetric systems, which are described by the COE ensemble. On the other hand, the CUE ensemble models chaotic cavities where time-reversal invariance has been broken, for instance, a two-dimensional degenerate electron gas in a diffusive periodic structure subject to a perpendicular magnetic field. For a chaotic quantum dot, i.e. a chain of length $L=1$, it is well known that the conductance is given by~\cite{Beenakker2} 
\begin{equation}
\langle g(L=1)\rangle = \frac{N}{2}+\left(1-\frac{2}{\beta}\right) \frac{1}{4},
\end{equation}
where $\beta=1$ for the COE case and $\beta=2$ for the CUE case. The difference $\delta g(1) = \langle g(1)\rangle_{\rm COE}-\langle g(1)\rangle_{\rm CUE} = - 1/4$ is called Weak Localization Correction (WLC) and can be explained semiclassically by the enhancement of the reflection probability due to constructive interference of time-reversed trajectories in time-reversal invariant systems~\cite{datta}. In the disordered wire, the WLC is observer in the metallic (ohmic) regime and is of slightly larger magnitude with $\delta \hat{g}(L) = -1/3$ independent of $L$. 

\begin{figure}[t]
 \centering
  \includegraphics[width=1\columnwidth]{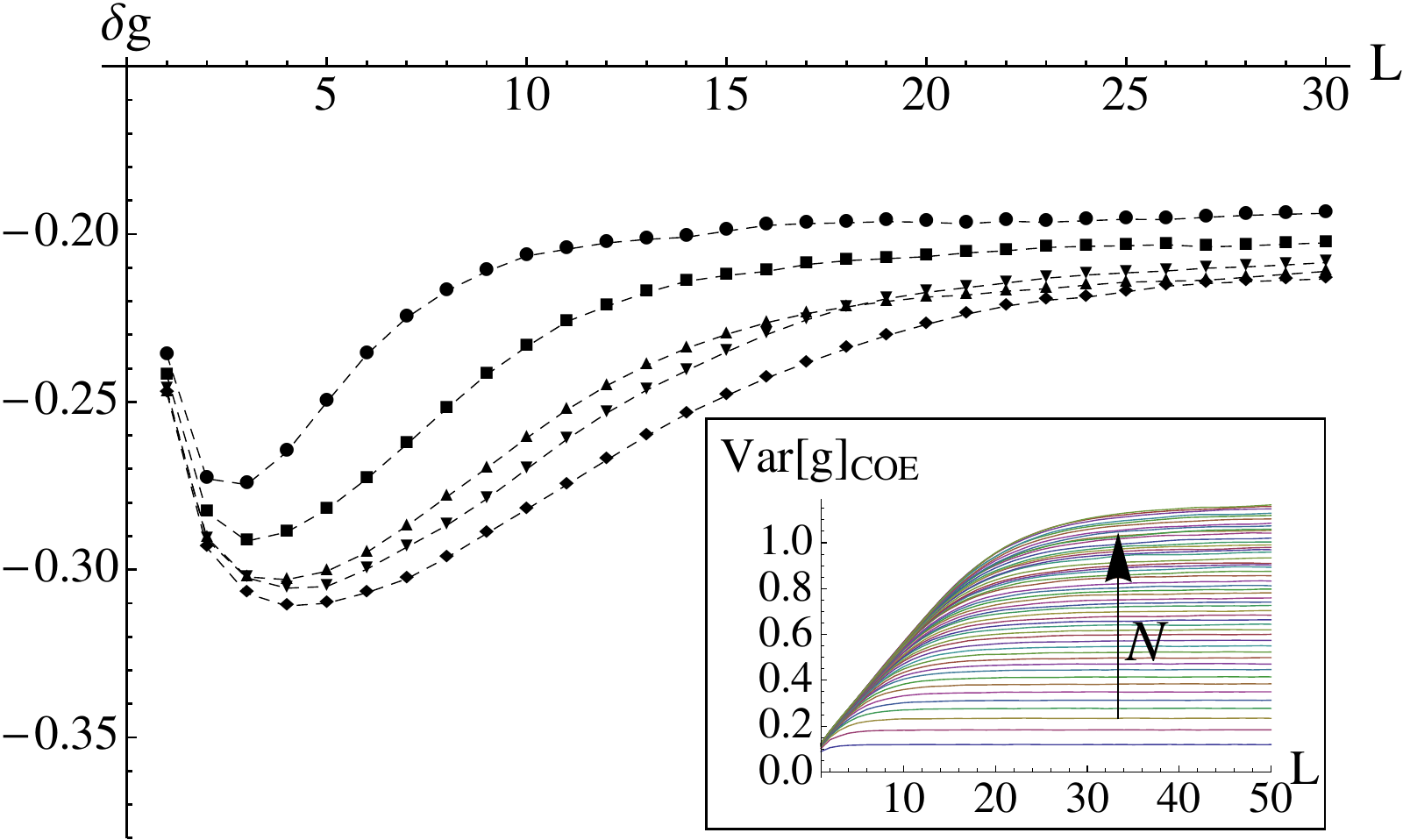}
 \caption{For the RMT model, plots of the WLC $\delta g(L)$ for $N=10,20,30,40,50$ using the same symbols as in Fig.~\ref{Fig3}. In the inset we plot the conductance variance ${\rm Var}[g(L)]$ for $N=1$ to $N=50$ for the COE case.}
 \label{Fig.wl} 
\end{figure}

Now we consider the WLC in a periodic chain.  It can be assumed that a weak magnetic field will not change the diffusion coefficient of a chain of strongly chaotic systems, thus we can extract the WLC in the periodic chain as a function of $L$ analogously to the quantum dot as $\delta g(L) = \langle g(L)\rangle_{\rm COE}-\langle g(L)\rangle_{\rm CUE}$. In Fig.~\ref{Fig.wl} we plot $\delta g(L)$, which for an $N$-dependent $L$-range is close to the weak localization correction of a disordered wire~\cite{BeenakkerRMP}. Although weak localization is usually associated to the metallic regime, here we see that in periodic systems it extends for large $L$ with a correction $\delta g(L) \approx -0.2$ that persists during the Bloch-ballistic regime. 

Recently, the WLC for a periodic system was studied~\cite{tian} assuming that the Ehrenfest time of the cavity is larger than the ergodic 
time~\cite{ergodic-time} of the (closed) cavity. We have in mind the opposite case, where the Ehrenfest time is smaller than the ergodic time ensuring that RMT is a good model for the periodic waveguide.

\subsection{Conductance Fluctuations}

\begin{figure}[t]%
 \centering
 \includegraphics[width=1\columnwidth]{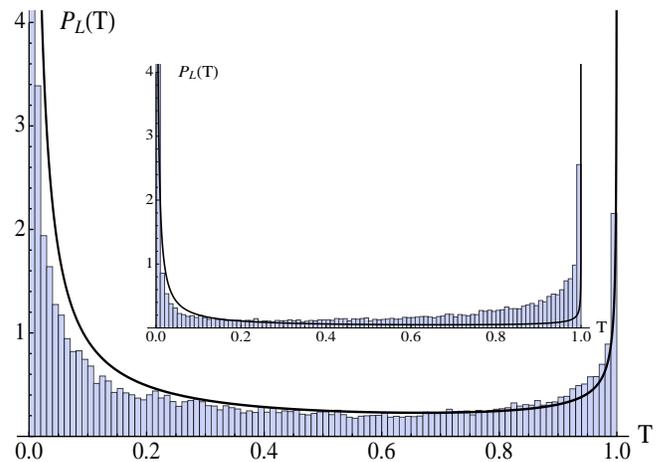}
 \caption
 {Transmission pdf $P_L(T)$ of a cosine periodic waveguide chain model with $N=50$ for $L=5$ in the diffusive regime. 
 The continuous line represents Dorokhov's distribution for a disordered wire $P_L(T)=\frac{N}{2L}\frac{1}{T\sqrt{1-T}}$. Histogram is computed with the same parameters than Fig.~\ref{fig-T_i}.  The inset shows the histogram of
 $P_L(T)$ for $L=25$, deep in the Bloch-ballistic regime and the continuous line is the corresponding Dorokhov's distribution.}
 \label{figdorokhov} 
\end{figure}

\begin{figure}[t]%
 \centering
 \includegraphics[width=0.48\columnwidth]{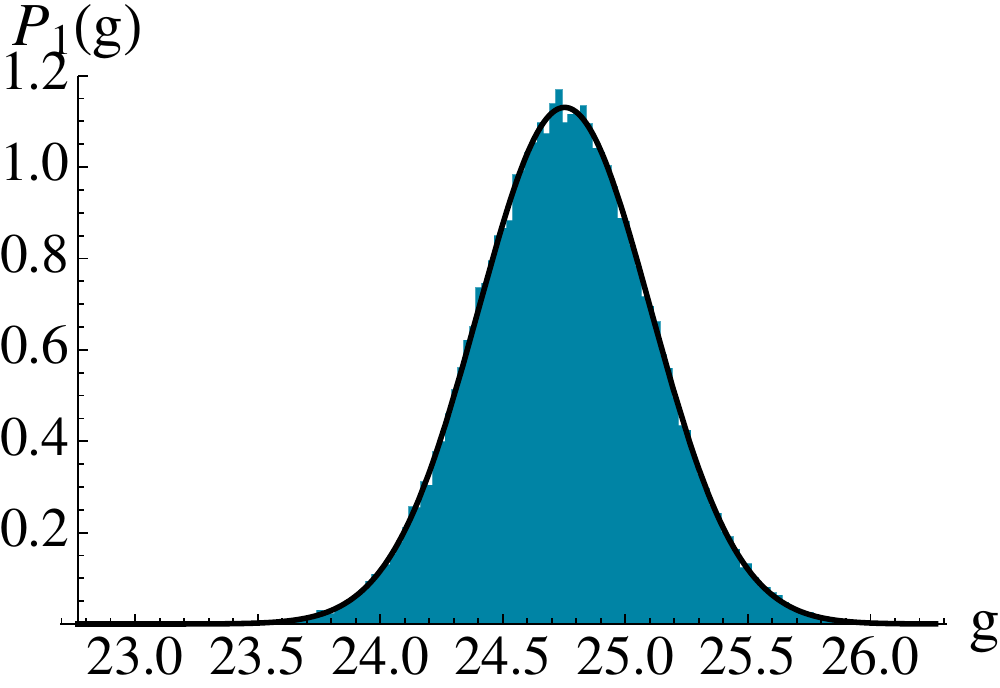}
 \includegraphics[width=0.48\columnwidth]{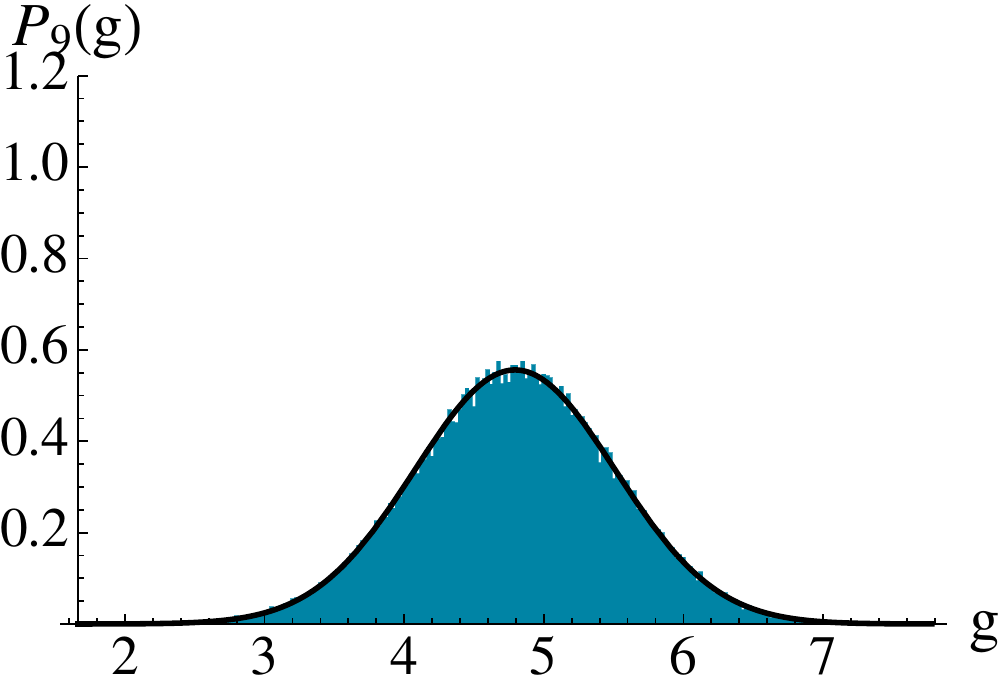}\\
 \vspace{0.1cm}
 \includegraphics[width=0.48\columnwidth]{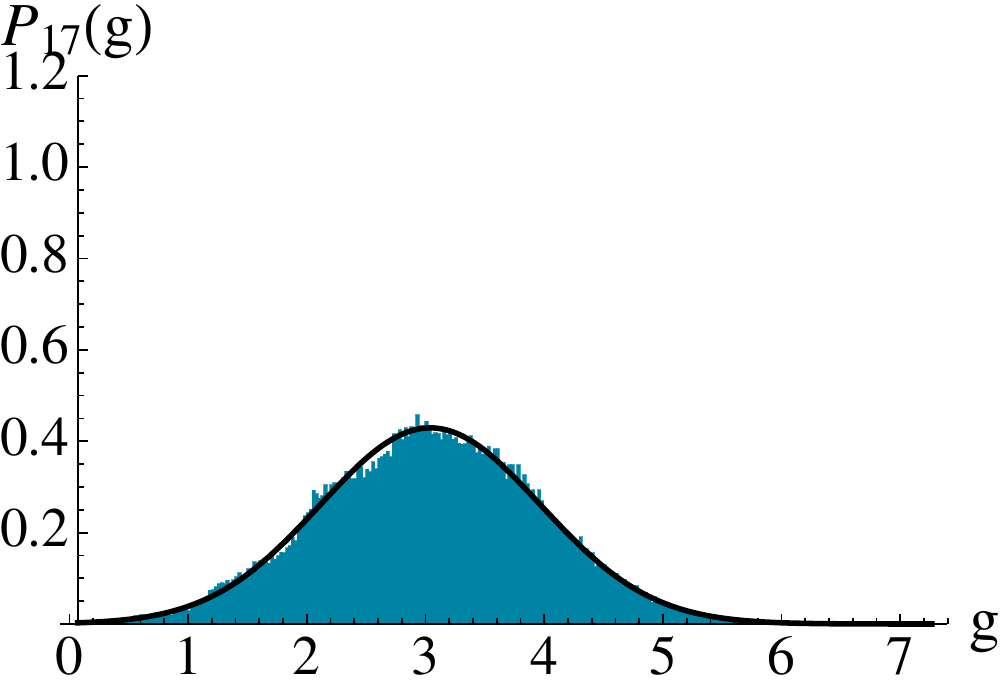}
 \includegraphics[width=0.48\columnwidth]{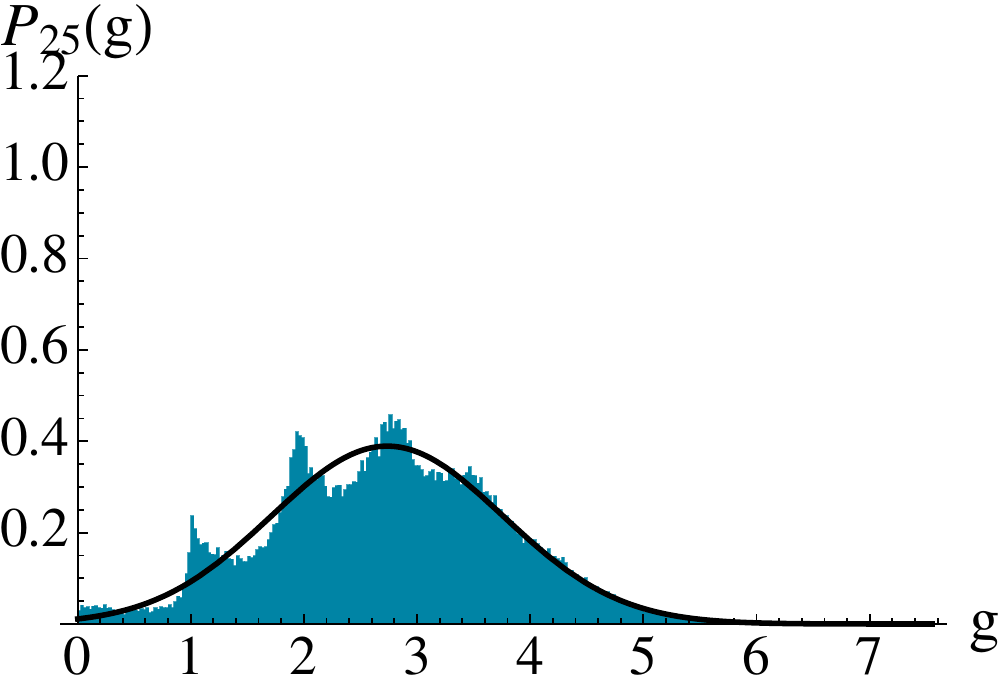}\\
  \vspace{0.1cm}
 \includegraphics[width=0.48\columnwidth]{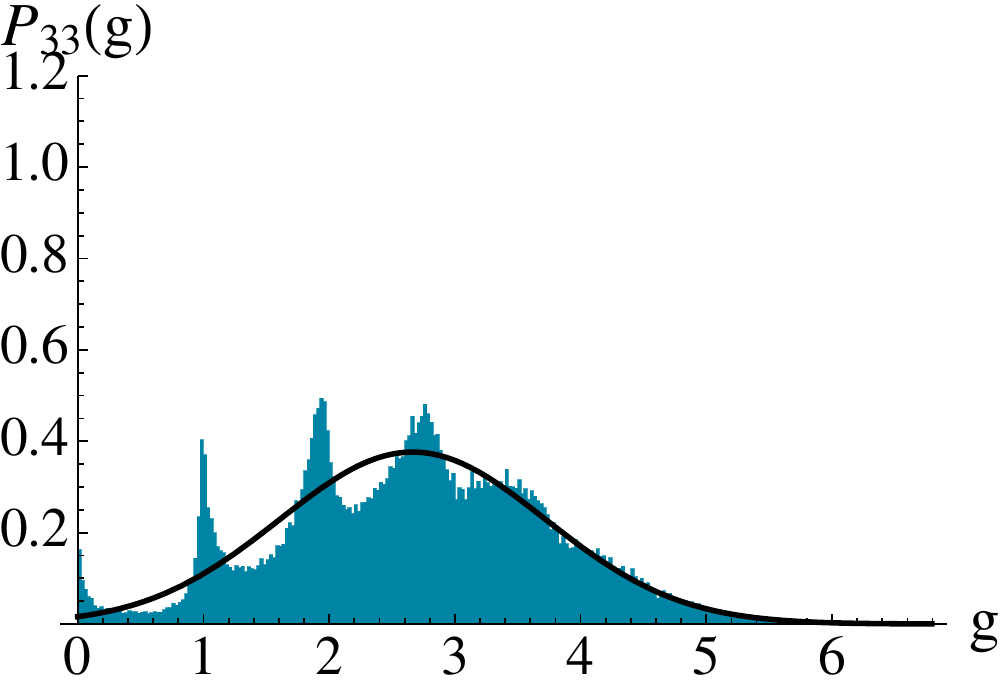}
 \includegraphics[width=0.48\columnwidth]{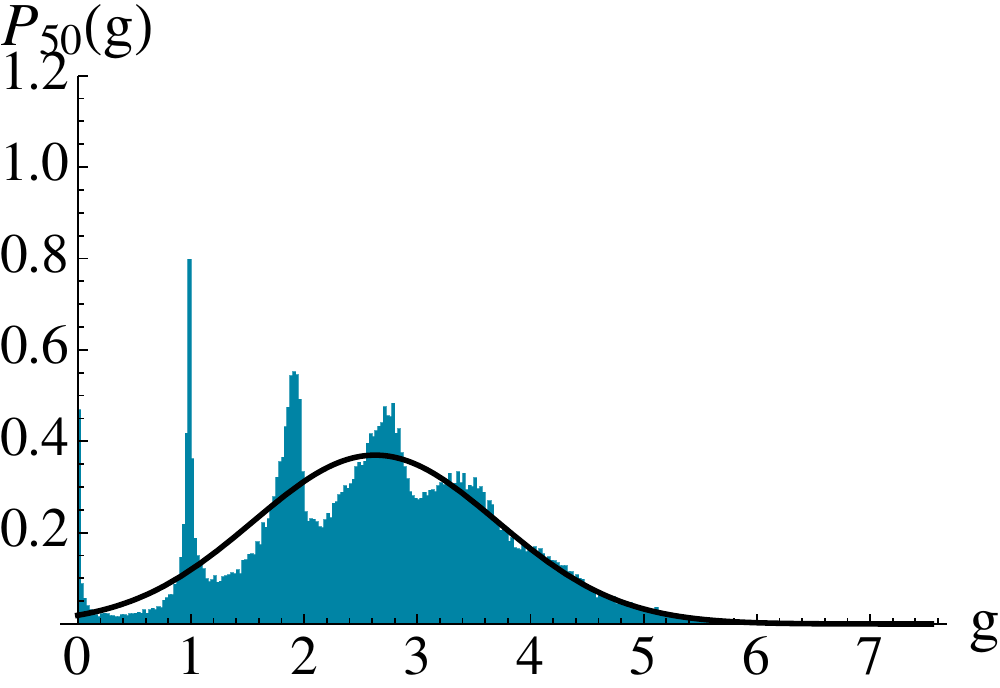}
 \caption
 {Conductance pdf $P_L(g)$ of a COE periodic chain model with $N=50$ for several values of $L$. We see that up to $L=9$ the distribution is clearly gaussian (black line) and at $L=17$ peaks start to develop. At $L=50$ the distribution has reached its long-chain stationary form consisting in a multimodal structure which arises from the asymptotic quasi-periodic behavior of $g$ around its average $\sim N_B$. Given an integer $n$ the peak just below $g=n$ is due to realizations with $N_B=n$. Note that for $N_B=1$ the peak is very narrow and its center close to one but for bigger $N_B$ the peaks start to spread and locate further away from its respective integer upper boundary. This issue is addressed in the main text. Note that the $x$-axis range covered changes in each plot.}
 \label{Fig.distrib} 
\end{figure}

In the previous discussion we noticed that finite periodic waveguides with chaotic cells possess a diffusive regime where $\langle g(L) \rangle$, including its WLC, are similar to those of the metallic regime in a disordered wire. We now address conductance fluctuations. In a disordered wire in the metallic regime, the conductance has a Gaussian distribution with ${\mathcal O}(1)$ variance, independent of the system length and of the disorder properties. The conductance distribution changes from this Gaussian form to a log-normal distribution as the size of the wire reaches the localization length. In our periodic chain, we also found that in the diffusive regime the conductance has Gaussian fluctuations [see Fig.~\ref{Fig.distrib}], however its variance ${\rm Var}[g(L)]=\beta L$ depends linearly on $L$ [see inset in Fig.~\ref{Fig.wl}] but with
a slope $\beta \approx 0.05$ sufficiently small to ensure that the average is representative of a typical value, e.g. we check that $1/\langle g(L) \rangle \approx \langle 1/g(L) \rangle$ in this regime. Let's consider the second order expansion 
\begin{equation}\label{g-moments-exp}
\frac{\langle g^{-1}(L)\rangle}{\langle g(L)\rangle^{-1}} = 1 - \mbox{Var}[g(L)]\, \langle g(L)\rangle^{-2}  + \mathcal{O}\left(\langle g(L)\rangle^{-3} \right) .
\end{equation}
Since $\langle g(L)\rangle = N/(L+1)$ we have that, deep in the ohmic regime, where $L\ll\sqrt{N}<N$, $\mbox{Var}[g(L)]\ll L$ and $ \mbox{Var}[g(L)]\, \langle g(L)\rangle^{-2} \ll 1$. Hence, keeping the dominant term in \eref{g-moments-exp} we obtain $\langle 1/g(L)\rangle \sim 1/\langle g(L)\rangle$.
Inset in Fig.~\ref{Fig.wl} shows that the variance grows linearly with $L$ for $L \lesssim \sqrt{N}$ and then reaches a constant value with ${\rm Var}[g(L)]\sim\mathcal{O}(\sqrt{N})$, signaling the diffusive to Bloch-ballistic transition.  From the RMT quantum dot description we know that  ${\rm Var}[g(L=1)]=1/8$.  On the other hand, the limit value ${\rm Var}[g(L\to\infty)]$ can be understood from the fluctuations of $N_B$ which are of order $\mathcal{O}(\sqrt{N})$~\cite{tobe}. 
The similarities and differences between the diffusive regime of periodic systems and disordered
systems is also illustrated in Fig.~\ref{figdorokhov} where we compare  histogram of the probability distribution function  $P_L(T)$ of transmission eigenvalues $T_i$ for the periodic cosine waveguide
with the Dorokhov's distribution for a disordered wire~\cite{dorokhov1982,BeenakkerRMP}. We can observe that qualitatively the distributions resembles each other but they differ quantitatively, for instance for the periodic waveguide there is a slightly increase in the population of transmission eigenvalues 
larger than $0.8$ and a corresponding decrease in the eigenvalues smaller than that. 
On the other hand, as illustrate the inset in Fig.~\ref{figdorokhov} deep in the Bloch-ballistic regime large transmission probabilities occurs much more often than
in the diffusive regime as expected. 

A remarkable feature that characterize the Bloch-ballistic regime is the conductance distribution multimodal shape observed in  Fig.~\ref{Fig.distrib}. In fact, as we have already mentioned, the conductance in the Bloch-ballistic regime is dominated by the first term in Eq.~(\ref{g}) that represent the sum $\sum_{i=1}^{N_B}T_i$ over the $N_B$ non-decaying  $T_i$. Therefore the fluctuations of $g(L)$ are related to the fluctuations of these $T_i$ and also to the fluctuations in their number $N_B$ which take only integer values. The sharp peaks of the conductance distribution $P_L(g)$ at $g=0$ and $g=1$ corresponds to realizations with $N_B=0$ and $N_B=1$, respectively. 
We see that peaks at larger value of $g$ becomes less sharp and also are not located at integer values of $g$. This is due to the repulsion between transmission coefficients which we have discussed in section \ref{cond}.  Owing to this repulsion, only one $T_i\sim 1$ and the rest are necessarily smaller, spreading in the [0,1] interval as shown in Fig.\ref{fig-T_i}, thus $g(L) = \sum_{i=1}^{N_B}T_i<N_B$. 

\vspace{0.2cm}

\section{conclusions}\label{conc}

In this work we have studied the propagation of waves in diffusive periodic quasi-one-dimensional systems, by numerically computing the conductance of a cosine-shaped waveguide and by employing a RMT model to describe the system. We have shown that wave propagation in such systems displays a diffusive regime for systems of length $L$ in the range $1 \ll L\ll \sqrt{N}$, where the conductance varies from ${\mathcal O}(N)$ to ${\mathcal O}(\sqrt{N})$. This should be compared with the diffusive regime of disordered systems which holds for $1 \ll L\ll N$, where the conductance varies from ${\mathcal O}(N)$ to ${\mathcal O}(1)$. The ohmic behavior of bulk disordered wires has been studied for a long time~\cite{tamura} and more recently was reported in surface disordered wires~\cite{saenz}. Here we have shown that this regime is also observed in periodic chains of cavities with diffusive classical dynamics.

On the other hand, for periodic waveguides with length $L \gg \sqrt{N}$, wave propagation acquires the ballistic character of the Bloch states of the associated unfolded infinite periodic system, with a constant average conductance $\langle g(L)\rangle$ which is close (and bounded by) $\langle N_B\rangle$. We found that there is a WLC in the conductance both in the ohmic and Bloch-ballistic regimes. In the former, we observe a value similar to disordered wires with $\delta g(L) \approx 0.3$, whereas in the latter the corrections is somewhat smaller with  $\delta g(L) \approx 0.2$. 

A difference we have observed between the ohmic regimes in disordered and periodic systems is in the conductance fluctuations. While the conductance variance ${\rm Var}[g(L)]$ is ${\mathcal O}(1)$ for the disordered wire, in diffusive periodic waveguides it grows linearly with $L$ up to $L\sim\sqrt{N}$ and then reaches a constant asymptotic value. The passage between these two regions signals the diffusive to Bloch-ballistic transition.   

\begin{acknowledgments}
FB and JZ thank ACT 127 and Conicyt for support. F.B. thanks Fondecyt Project No. 1110144.
F.B and V.P thank ECOS project C09E07 and Anr-Conicyt project 38.
\end{acknowledgments} 

\nopagebreak


\begin{thebibliography}{99}
\bibitem{ergodic-time} The ergodic time is the characteristic time needed for an initial density of classical particles to relax to the uniform invariant density.
\end{thebibliography}
\end{document}